\documentclass[conference]{IEEEtran}
\IEEEoverridecommandlockouts

\usepackage{amsmath,amsfonts,graphicx}
\usepackage{amssymb,textcomp,mathtools}
\usepackage{bm,upgreek,algorithm,hyperref}
\usepackage{multirow,booktabs,hhline,array}
\usepackage{cite,url,makecell,setspace}
\usepackage{xcolor}
\pdfoutput=1

\renewcommand{\vec}[1]{\bm{\mathrm{#1}}}

\title{SPMamba: State-space model is all you need in speech separation
\thanks{$^*$Equal contribution. $^\dagger$Corresponding author}}
\author{\IEEEauthorblockN{Kai~Li$^{1,*}$, Guo Chen$^{1,*}$, Runxuan Yang$^{1}$, Xiaolin Hu$^{1,2,\dagger}$ \\}
\IEEEauthorblockA{$^1$Department of Computer Science and Technology, BNRist, \\THBI, IDG/McGovern Institute for Brain Research, Tsinghua University, Beijing, China \\
$^2$Chinese Institute for Brain Research (CIBR), Beijing, China \\
tsinghua.kaili@gmail.com, cg22@mails.tsinghua.edu.cn, yangrx20@mails.tsinghua.edu.cn, xlhu@tsinghua.edu.cn}
}

\begin{document}
\maketitle

\begin{abstract}
Existing CNN-based speech separation models face local receptive field limitations and cannot effectively capture long time dependencies. Although LSTM and Transformer-based speech separation models can avoid this problem, their high complexity makes them face the challenge of computational resources and inference efficiency when dealing with long audio. To address this challenge, we introduce an innovative speech separation method called SPMamba. This model builds upon the robust TF-GridNet architecture, replacing its traditional BLSTM modules with bidirectional Mamba modules. These modules effectively model the spatiotemporal relationships between the time and frequency dimensions, allowing SPMamba to capture long-range dependencies with linear computational complexity. Specifically, the bidirectional processing within the Mamba modules enables the model to utilize both past and future contextual information, thereby enhancing separation performance. Extensive experiments conducted on public datasets, including WSJ0-2Mix, WHAM!, and Libri2Mix, as well as the newly constructed Echo2Mix dataset, demonstrated that SPMamba significantly outperformed existing state-of-the-art models, achieving superior results while also reducing computational complexity. These findings highlighted the effectiveness of SPMamba in tackling the intricate challenges of speech separation in complex environments. 
\end{abstract}
\section{Introduction}
\label{sec:introduction}
In recent years, deep learning-based speech separation models have made significant progress. Models based on Convolutional Neural Networks (CNNs)\cite{luo2019conv, tzinis2020sudo, li2022efficient, hu2021speech}, Recurrent Neural Networks (RNNs)\cite{li22d_interspeech,li2023design,luo2020dual}, and Transformer \cite{chen2020dual, subakan2021attention, yang2022tfpsnet} have shown outstanding performance in this task. However, each of these three categories of models presents certain limitations in speech separation. Constrained by their local receptive fields, CNN-based models \cite{luo2019conv, tzinis2020sudo, li2022efficient, hu2021speech} struggle to capture the complete context of audio signals, which can impair their separation performance. While LSTM has been used to address the issue of long-term dependencies, it \cite{li22d_interspeech,li2023design,luo2020dual,chen2020dual,wang2023tf} still faces challenges in effectively capturing distant dependencies due to its short-term memory. The Transformer architecture is renowned for its exceptional ability to model long-range dependencies in large-scale models, yet its self-attention mechanism exhibits quadratic complexity of the sequence length, resulting in high computational costs in real-time applications \cite{chen2020dual, subakan2021attention, yang2022tfpsnet}. 

Recently, the development of State Space Models (SSMs) \cite{gu2021efficiently,gu2021combining} has shown potential in overcoming these limitations. By incorporating enhanced selection mechanisms, SSMs effectively address the challenges encountered by traditional models. The introduction of weight modulation during propagation significantly expands the effective receptive field. It establishes long-range dependencies with linear computational complexity, making them particularly well-suited for tasks requiring efficient processing of long sequences. Building on the foundation of classical SSM research, modern SSMs, exemplified by models like Mamba, have demonstrated remarkable effectiveness across various domains, including natural language processing \cite{pioro2024moe,yang2024clinicalmamba} and visual tasks \cite{liu2024vmamba,liu2024swin}. However, in speech separation, the potential of SSMs as a model design approach remains largely untapped.

\begin{figure*}[h]
	\small
	\centering
	\includegraphics[width=2.0\columnwidth]{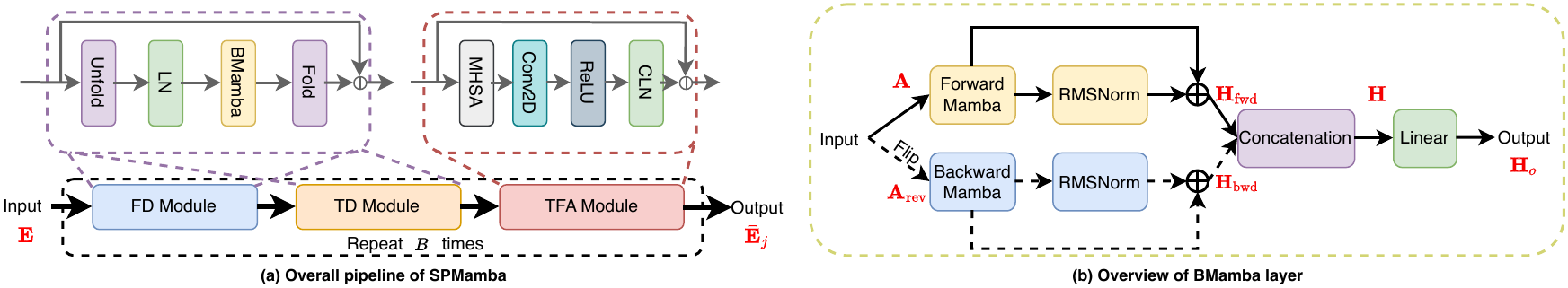}
	\caption{An overview of the proposed SPMamba model and BMamba layer. The BMamba layer processes both forward and backward audio sequences. ``LN" denotes the layer normalization, and ``CLN" denotes the cumulative layer normalization.}
	\label{fig:spmamba}
\end{figure*}

Leveraging the transformative potential of SSMs in capturing long-range dependencies with linear computational complexity, we introduce an innovative architecture for speech separation called SPMamba. 
This architecture ingeniously integrates the essence of SSMs into the realm of audio processing, specifically targeting the challenges of speech separation. 
SPMamba is built upon the robust framework of TF-GridNet \cite{wang2023tf}, renowned for effectively handling temporal and frequency dimensions in audio signals. By replacing the BLSTM modules of TF-GridNet with bidirectional Mamba modules (see Fig. \ref{fig:spmamba}(a)), we expect that the model's ability to process audio sequences in expansive contextual scenarios will be enhanced. This substitution mitigates the computational inefficiencies inherent in RNN-based approaches.

We conducted extensive experiments on three public datasets (WSJ0-2Mix \cite{hershey2016deep}, WHAM! \cite{wichern2019wham}, and Libri2Mix \cite{cosentino2020librimix}) and noise- and reverberation-augmented dataset (Echo2Mix) constructed using LibriSpace \cite{LibriSpace2024}. 
Experiments showed that SPMamba achieved performance metrics on par with traditional separation models on public datasets (WSJ0-2Mix, WHAM!, and Libri2Mix) while exhibiting considerably lower parameter counts and computational demands than SOTA separation models. On the Echo2Mix dataset, SPMamba achieved a substantial 2.42 dB improvement in SI-SNRi compared to TF-GridNet. This improvement represents a fundamental leap in the quality of separation achieved by integrating the SSM into the frequency and time modeling modules.

\section{Revisiting Mamba Model}
\label{sec:mamba}
The Mamba method \cite{gu2023mamba} introduces a novel approach by employing Selective SSM that combines the strengths of both CNNs and RNNs, and addresses their limitations through a selection mechanism that incorporates input-dependent dynamics. This Selective SSM technique enables the model to selectively focus on or ignore parts of the input sequence, which is crucial for separating overlapping speech signals.

Structured SSMs \cite{gu2021efficiently,gu2021combining} and Mamba are inspired by the continuous system, which maps an input function $x(t)\in \mathbb{R}$ to an output $\dot{y}(t)\in \mathbb{R}$ through a higher-dimensional latent state $h(t)\in \mathbb{R}^{N}$ cross all continuous time steps $T$, where $N$ denotes the number of feature channels, as illustrated in the following equations:
\begin{equation}
    \dot{h}(t) = \mathbf{A}h(t) + \mathbf{B}x(t), y(t) = \mathbf{C}^{\top}h(t), 
\label{eq:1}
\end{equation}
where $\mathbf{A}\in \mathbb{R}^{N\times N}$ denotes the evolution parameter, $\mathbf{B}\in \mathbb{R}^{N\times 1}$ and $\mathbf{C}\in \mathbb{R}^{N\times 1}$ denote the projection parameters. 

Structured SSMs originate from continuous time mapping (see Equation \ref{eq:1}) and, therefore, cannot be directly applied to discrete sequences. By using the Zero Order Holding (ZOH) method (see Equation \ref{eq:2}), the Structured SSM completes the discretization process, in which the values are sampled only according to the input time step $\Delta$:
\begin{equation}
\bar{\mathbf{A}} = \exp(\Delta \mathbf{A}), \bar{\mathbf{B}} = (\Delta \mathbf{A})^{-1} (\exp(\Delta \mathbf{A}) - I) \cdot \Delta \mathbf{B}.
\label{eq:2}
\end{equation}
\begin{equation}
    h(t) = \bar{\mathbf{A}}h(t-1) + \bar{\mathbf{B}}x(t), y(t) = \mathbf{C}^{\top}h(t), 
\label{eq:3}
\end{equation}
where the parameters ($\bar{\mathbf{A}}$, $\bar{\mathbf{B}}$, $\mathbf{C}$) can be varied over time in a form known as selective state space modeling (SSM), analogous to a gating mechanism in RNNs. This mechanism indicates that the model is able to selectively focus on or ignore input features at each time step, thus enhancing the information capacity of the model \cite{gu2023mamba}.

\section{Method}
\label{sec:model}
\subsection{Overall Pipeline}
\label{sec:overall}

The overall separation architecture employed by SPMamba aligns with time-frequency domain separation methods \cite{wang2023tf}. Specifically, when the input mixed signal $\mathbf{Q} \in \mathbb{R}^{1 \times L}$ is provided, we utilize the Short-Time Fourier Transform (STFT) as the encoder to encode $\mathbf{Q}$ into an input representation, denoted as $\mathbf{E} = \text{STFT}(\mathbf{Q}) \in \mathbb{R}^{2\times F \times T_o}$, where $L$, $F$ and $T_o$ represent the audio length, the frequency and time dimensions, respectively. Subsequently, the separator estimates $J$ output representations $\Bar{\mathbf{E}}_j$, which are then decoded by the audio decoder into $\Bar{\mathbf{X}}_j = \text{iSTFT}(\Bar{\mathbf{E}}_j) \in \mathbb{R}^{1 \times L}$, where $1 \leq j \leq J$, as shown in Fig. \ref{fig:spmamba}(a). Following the separator design of TF-GridNet \cite{wang2023tf}, SPMamba directly maps the output signals back to the time domain via inverse STFT (iSTFT) rather than performing mask operations.

\subsection{SPMamba Model}\label{sec:SPMamba}

The SPMamba adopts TF-GridNet \cite{wang2023tf} as its backbone network, which represents the SOTA speech separation model. To further improve the model's efficiency, we replace the BLSTM network with a bidirectional Mamba network in the separator. Next, we will elaborate on the structural design and implementation details of the SPMamba model. SPMamba consists of three components (Fig. \ref{fig:spmamba}(a)): 1) frequency-domain (\textbf{FD}) module; 2) time-domain (\textbf{TD}) module; and 3) time-frequency attention (\textbf{TFA}) module.

\textbf{FD Module.} In this module, we treat the input tensor as a series of independent frequency sequences and employ a BMamba layer to capture the complex relationships within each frequency. First, we unfold the input tensor using a kernel of size $K$ and stride $S$ to enhance the local spectral context.
Next, we apply layer normalization along the channel dimension of the unfolded tensor, followed by a BMamba layer (Section \ref{sec:bmamba}) with $H=256$ hidden units in each direction to model the intra-frame frequency information. To recover the original dimensions, we use a 1D Transpose convolutional layer with kernel size $K=8$, stride $S=1$, input channel $2H$, and output channel $C=128$ to process the hidden embeddings of the BMamba layer. Finally, we add the input tensor to the output tensor via a residual connection to facilitate gradient flow and learning. 

\textbf{TD Module.} In this module, the procedure closely resembles the frequency-domain module. The key distinction lies in interpreting the input tensor, which is treated as F-independent sequences here, each with a length of T. Within this module, a BMamba layer is employed to capture the temporal information present within each sub-band.

\textbf{TFA Module.} Like TF-GridNet, this module leverages frame-level embeddings derived from the time-frequency representations within each frame of the output tensor generated by the time-domain module. It employs whole-sequence self-attention on these frame embeddings to capture long-range global information. The concatenated attention outputs undergo further processing to obtain the output tensor, which is fed into the next SPMamba block.

\subsection{BMamba}\label{sec:bmamba}

While the Mamba exhibits unique features, its causal processing of the input data limits it to capture only historical information about the data. 
However, our focus is on non-causal speech separation. To overcome this limitation, an intuitive solution is to mimic the processing of BLSTM by scanning speech frames along both forward and backward directions, thus enabling the model to combine current and historical features. Fig. \ref{fig:spmamba}(b) shows the detailed structure of the BMamba layer.

Specifically, we process the input audio feature $\mathbf{A}\in \mathbb{R}^{N \times K}$ from both the forward and backward directions, where $N$ and $K$ denote the channels and length of features, respectively. To achieve this bidirectional processing, we first apply a $\text{Mamba}_f(\cdot)$ to capture the temporal dependencies of the feature sequence. This yields a hidden representation $\mathbf{H}_\text{fwd} = \text{Mamba}_f(\mathbf{A})\in \mathbb{R}^{N \times K}$. Next, we reverse the features sequence and input it into $\text{Mamba}_b(\cdot)$, which operates from the end to the start of the sequence, providing another hidden representation $\mathbf{H}_\text{bwd} = \text{Mamba}_b(\mathbf{A}_{\text{rev}})\in \mathbb{R}^{N \times K}$, where $\mathbf{A}_{\text{rev}}$ represents the reversed input features. Then, the outputs from both the forward and backward passes are combined through concatenation to produce the bidirectional features $\mathbf{H}\in \mathbb{R}^{2N \times K}$:
\begin{equation}
    \mathbf{H} = \text{Concat}(\mathbf{H}_\text{fwd}, \mathbf{H}_\text{bwd}),
\end{equation}
where $\text{Concat}(\cdot, \cdot)$ represents the concatenation operation used to merge the forward and backward hidden representations. Finally, we use the linear layer to compress the channels of $\mathbf{H}$ to obtain the final output $\mathbf{H}_o\in \mathbb{R}^{N \times K}$.  This bidirectional processing ensures that both past and future contexts are considered when processing the audio features, leading to a more robust and context-aware representation.

\section{Experiment configurations}
\label{sec:config}
\begin{table*}[]
\footnotesize
\centering
\caption{Quantitative comparison of SPMamba with other existing models.
$*$ denotes the results of TF-GridNet using espnet\cite{watanabe2018espnet}.}
\begin{tabular}{ccccccccccc}
\toprule
\multirow{2}{*}{Model} & \multicolumn{2}{c}{WSJ0-2Mix} & \multicolumn{2}{c}{WHAM!}     & \multicolumn{2}{c}{Libri2Mix} & \multicolumn{2}{c}{Echo2Mix}    & \multirow{2}{*}{Params(M)} & \multirow{2}{*}{Macs (G/s)} \\ \cmidrule(r){2-3} \cmidrule(r){4-5} \cmidrule(r){6-7} \cmidrule(r){8-9}
                       & SDRi          & SI-SNRi       & SDRi          & SI-SNRi       & SDRi          & SI-SNRi       & SDRi           & SI-SNRi        &                            &                             \\ \midrule
Conv-TasNet            & 15.6          & 15.3          & 13.0             & 12.7          & 12.7          & 12.2          & 7.7           & 6.9           & 5.62                       & 10.23                       \\
DualPathRNN            & 19.0          & 18.8          & 14.1          & 13.7          & 16.6          & 16.1          & 5.9           & 5.1           & 2.72                       & 85.32                       \\
SudoRM-RF              & 19.1          & 18.9          & 14.1          & 13.7          & 14.4          & 14.0          & 7.7           & 6.8           & 2.72                       & \textbf{4.60}               \\
A-FRCNN                & 18.6          & 18.3          & 14.8          & 14.5          & 17.2          & 16.7          & 9.6           & 8.8           & 6.13                       & 81.20                       \\
TDANet                 & 18.7          & 18.5          & 15.4          & 15.2          & 17.9          & 17.4          & 10.1          & 9.2           & \textbf{2.33}              & 9.13                        \\
BSRNN                  & 20.9          & 20.7          & 16.5          & 16.2          & 15.7          & 15.2          & 12.8          & 12.2          & 25.97                      & 98.69                       \\
TF-GridNet             & \textbf{23.1} & \textbf{22.8} & 17.2*         & 16.9*         & 20.1*         & 19.8*         & 13.7          & 12.8          & 14.43                      & 445.56                      \\
SPMamba (Ours)         & 22.7          & 22.5          & \textbf{17.6} & \textbf{17.4} & \textbf{20.4} & \textbf{19.9} & \textbf{16.1} & \textbf{15.3} & 6.14                       & 238.69                     \\ \bottomrule
\end{tabular}
\label{tab:com}
\end{table*}

\subsection{Public Benchmark Datasets}

Following previous methods \cite{luo2019conv,li2022efficient,wang2023tf}, we trained and tested our proposed SPMamba on the public benchmark datasets WSJ0-2Mix\cite{hershey2016deep}, WHAM!\cite{wichern2019wham}, and Libri2Mix\cite{cosentino2020librimix}. The network was trained on these datasets using 4-second long segments at an 8 kHz sampling rate. WSJ0-2Mix\cite{hershey2016deep} is a clean audio dataset.
WHAM!\cite{wichern2019wham} is a noisy version of WSJ0-2Mix
, thereby increasing the separation challenge. Libri2Mix\cite{cosentino2020librimix} selects speech from a subset of LibriSpeech and mixes it at SNRs ranging from -25 to -33 dB, consistent with previous studies \cite{li2022efficient,hu2021speech}.

\subsection{Echo2Mix}
Although existing public datasets \cite{hershey2016deep, cosentino2020librimix, wichern2019wham} are widely used in speech separation, the reverberant environments in these datasets are usually simulated using regular geometries that do not accurately model the obstacles and materials in the scene compared to the acoustic characteristics of real-world environments. This simplification leads to significant deviations between the datasets and the real environment. We then constructed a multi-speaker speech separation dataset with noise and reverberation, which poses greater challenges than existing public datasets. For the speaker audio segments in the dataset, we selected publicly available data from the Librispeech dataset \cite{panayotov2015librispeech}, specifically utilizing the Librispeech-360. The noise segments were sourced from the noise provided by the WHAM! dataset \cite{wichern2019wham} and the sound effects from the DnR dataset \cite{petermann2022cocktail}. The background music segments were drawn from the cleaned music portion of the DnR dataset \cite{petermann2022cocktail}.

To create realistic synthetic mixtures, we set the length of each mixture to 60 seconds with a sampling rate of 16 kHz. Two segments from the same speaker did not overlap, although overlaps between foreground and background sound effects were allowed. We used Loudness Units Full Scale (LUFS) \cite{grimm2010toward} to adjust the relative levels, setting the music at $-24$ LUFS, speech at $-17$ LUFS, and sound effects at $-21$ LUFS. We uniformly sampled an average LUFS value within a $±2.0$ range around the target LUFS for each mixture. For reverberation processing, we simulated spatial reverberation using LibriSpace \cite{LibriSpace2024} through convolution with impulse responses, achieving effects close to real-world scenarios. Ultimately, we constructed a dataset\footnote{\url{https://drive.google.com/file/d/1nJ9ujAbf4LxXEFzFwEpr9CwJOeNHghw0/view}} comprising 57 hours of training data, 8 hours of validation data, and 3 hours of test data, which was used to evaluate the performance of various models.

\subsection{Model Configurations}

For the STFT and iSTFT, we employed a 512-point Hann window with a hop size of 128 points. A 512-point Fourier transform was applied to extract a 129-dimensional complex spectrum for each frame. We used $B = 6$ blocks. Each BMamba layer comprised two Mamba components, with a hidden layer dimension of 128. RMSNorm \cite{zhang2019root} was utilized to normalize the output of the Mamba components.

\subsection{Training and Evaluation}

During the training process, we randomly selected 4s mixed audio segments for training. We chose negative SNR as the objective function for the PIT training method \cite{yu2017permutation}. The Adam optimizer \cite{kingma2014adam} was employed with an initial learning rate of 0.001, and the learning rate was halved if the validation loss did not improve within 10 epochs. The maximum value for gradient clipping was set to 5. The model was trained until no best validation model was found for 20 consecutive epochs. For evaluation, SI-SNRi \cite{le2019sdr} and SDRi \cite{vincent2006performance} were used as metrics. The number of parameters and MACs were calculated using ptflops \cite{ptflops}. The MACs were calculated by processing one audio with 1s in length and 16 kHz in sample rate. All experiments were conducted on a server with 10 $\times$ GeForce RTX 4090s. The source code for SPMamba is publicly accessible at \url{https://github.com/JusperLee/SPMamba}.

\section{Results}
\label{sec:result}
\subsection{Comparison with State-of-the-art Models}
We compared the SPMamba model with existing separation models on the benchmark datasets (WSJ0-2Mix, WHAM!, and Libri2Mix) and the newly constructed EchoMix dataset, as shown in Table \ref{tab:com}. The experimental results demonstrated that the proposed SPMamba method achieved the SOTA performance across multiple metrics. On the more complex Echo2Mix dataset, SPMamba achieved an SDRi of 16.1 dB and an SI-SNRi of 15.3 dB, setting the current state-of-the-art (SOTA) performance and significantly improving over other traditional methods. We provide several samples\footnote{\url{https://cslikai.cn/SPMamba/}} from the Echo2Mix dataset and the separated results of SPMamba and TF-GridNet for qualitative comparison. Additionally, as indicated in Table \ref{tab:com}, SPMamba's computational load was only half that of TF-GridNet in terms of computational complexity and parameters. 

\begin{figure}[h]
	\small
	\centering
	\includegraphics[width=0.7\columnwidth]{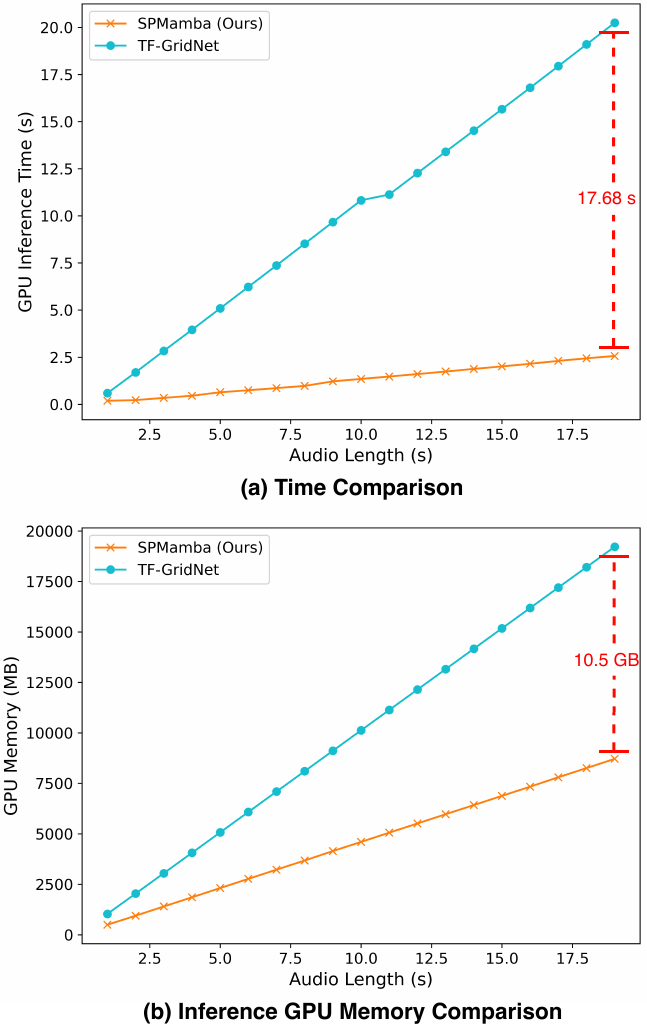}
	\caption{Efficiency comparisons between SPMamba and TF-GridNet.}
	\label{fig:efficient}
 \vspace{-10pt}
\end{figure}

We evaluated the complexity of SPMamba against TF-GridNet with input audio lengths ranging from 1 to 19 seconds. Fig. \ref{fig:efficient} shows that SPMamba significantly outperformed TF-GridNet regarding GPU memory usage and inference time, demonstrating the efficiency of SPMamba in long speech separation tasks.

\subsection{Impact of Integrating BMamba Layers}
Table \ref{tab:bmamba} illustrates the impact of integrating BMamba layers into the FD and TD modules of the baseline on the Echo2Mix dataset. The baseline model employed the TF-GridNet network architecture. The experimental results indicated that when the standard BLSTM was replaced with BMamba layers in the FD and TD modules, SDRi performance gains of 0.5 dB and 2.0 dB were achieved, respectively. When all BLSTM layers in both the FD and TD modules were replaced with BMamba layers, the SDRi performance improvement reached 2.4 dB. 
\begin{table}[h]
\footnotesize
\centering
\caption{Impact of integrating BMamba layers into different modules}
\begin{tabular}{cccc}
\toprule
FD Module                 & TD Module                 & SDRi & SI-SNRi \\
\midrule
$\times$                  & $\times$                  & 13.7 & 12.8    \\
\checkmark & $\times$                  & 14.2 & 13.4    \\
$\times$                  & \checkmark & 15.7 & 14.8    \\
\checkmark & \checkmark & 16.1 & 15.3    \\
\bottomrule
\end{tabular}
\label{tab:bmamba}
\vspace{-15pt}
\end{table}

\section{Conclusion}
\label{sec:conclusion}
We introduce a simple-yet-effective speech separation framework, SPMamba. Motivate by the need to capture long-range dependencies in audio sequences efficiently, we propose leveraging the strengths of bidirectional Mamba modules to replace traditional BLSTM components within the TF-GridNet architecture. By integrating these modules, SPMamba effectively improved the contextual understanding of audio signals, leading to superior performance in speech separation tasks. Extensive experiments demonstrated that SPMamba achieved remarkable results while with significantly lower computational complexity.

\bibliographystyle{IEEEtran}
\bibliography{refs}

\end{document}